\documentclass[11pt,twoside]{article}
\usepackage{asp2010}

\resetcounters

\begin{document}

\title{Thermohaline Instabilities Induced by Heavy Element Accretion onto White Dwarfs: Consequences on the Derived Accretion Rates.}
\author{M. Deal,$^{1}$ 
        S. Vauclair,$^{2,3}$ and
        G. Vauclair,$^{2,3}$ }
\affil{$^1$ Universit\'e de Toulouse, Toulouse, France}
\affil{$^2$ Universit\'e de Toulouse, UPS-OMP, IRAP, Toulouse, France}
\affil{$^3$ CNRS, IRAP, 14 avenue E. Belin, 31400, Toulouse}

\begin{abstract}
Heavy elements are observed in the atmospheres of many DA and DB white dwarfs, and their presence is attributed to the accretion of matter coming from debris disks. 
Several authors have deduced accretion rates from the observed abundances, taking into account the mixing induced by the convective zones and the gravitational settling. 
The obtained values are different for DA and DB white dwarfs. Here we show that an important process was forgotten in all these computations: thermohaline mixing, induced 
by the inverse $\mu$-gradient built during the accretion process. Taking this mixing into account leads to an increase of the derived accretion rates, specially for DA
 white dwarfs, and modifies the conclusions.
\end{abstract}

\section{Introduction: heavy elements accretion onto white dwarfs}

An increasing number of DA and DB white dwarfs observed with Spitzer show infrared excess due to circumstellar disks (\citet{farihi11}; \citet{xu12};
 \citet{girven12}; \citet{brinkworth12}). In the mean  time, high-resolution spectroscopy reveals the presence of heavy elements in
 the spectra of  many white dwarfs (reclassified as DAZ and DBZ) with effective temperatures below 25000~K  (\citet{zuckerman07}; \citet{klein10a}, \citet{klein10b}; 
 \citet{vennes10}; \citet{zuckerman11};  \citet{dufour12}). This was
 unexpected in such stars because heavy elements rapidly diffuse in the deep layers as a result of the very short gravitational settling time scales. The radiative
 accelerations which can balance gravity and support some heavy elements in hot white dwarfs (\citet{vauclair79}; \citet{chayer95}) 
is no longer operating in such cooler white dwarfs. These observations support the idea  that heavy elements are presently being accreted from circumstellar debris disks. 
The tidal disruption of asteroid-like bodies may be the source of the disk material polluting the white dwarf atmospheres (\citet{jura03}).
The determination of  the abundances of the heavy elements present in white dwarfs atmospheres may lead to an evaluation of the accretion rates, with the help  of stellar
 models. It is then possible to determine the mass and composition of the body whose disruption is at the origin of the pollution. Up to now,  such estimates have been 
obtained with the assumption that the accreting material is mixed into the outer convective zone and diffuses at its bottom due to gravitational settling. The derived 
accretion rates differ by several orders of magnitude between DAZ and DBZ (\citet{farihi12}), which is difficult to understand and does not seem to have 
any physical meaning. 
Here we show that these derived accretion rates are not correct, because a fundamental physical process has been forgotten in all these previous studies: thermohaline 
convection due to the accretion-induced inverse $\mu$-gradient. As a consequence, the mixed zone is much deeper than the classical convective zone, and the accretion rates 
are larger than those presently determined. We show here an example of computations of  this effect for a DA white dwarf. Due to the initial $\mu$ value, larger for helium 
than for hydrogen, and to the different internal structure, thermohaline mixing is less efficient in DB than in DA white dwarfs, for the same amount of accreted matter. This
 will be discussed in a forthcoming paper. We expect that this may reconcile the accretion rates needed to account for the observations of DAZ and DBZ white dwarfs.

\section{Thermohaline convection in stars}

Thermohaline convection, well known in oceanography, is now recognized as a major mixing process which  occurs in stars in the presence of an inverse $\mu$-gradient
 associated with a stable temperature gradient, so that the medium as a whole is dynamically stable. If a blob begins to move down, the heat exchange with its surroundings 
occurs more rapidly than the particle exchange so that the blob goes on falling, thereby triggering  the instability (see \citet{vauclair04}, 
\citet{vauclair12},  and references therein). 

Various situations may lead to thermohaline convection in stellar interiors. One  of them is the accretion of heavy matter onto the star, which may be due to planetary
 material (\citet{vauclair04}, \citet{garaud11},  \citet{theado12}) or to an evolved companion in a binary system (e.g. 
\citet{stancliffe07},  \citet{thompson08}). 
Thermohaline convection develops if:
\begin{equation}
$$1$\leq$ $R_{0}$ $\leq$ $\frac{1}{\tau}$$$
\end{equation}
where:
\begin{equation}
$$ $R_{0}$ = $\frac{\nabla_{ad}-\nabla}{|\nabla_{\mu}|}$$$
\end{equation}
and $\tau$ is the inverse Lewis number,
equal to the ratio of the particle diffusivity to the thermal diffusivity. Here $\nabla_{ad}$ and $\nabla$ are the usual
temperature gradients and $\nabla_{\mu}$ stands for $d \ln \mu/d \ln P$.

Thermohaline convection must clearly occur at the bottom of the outer convective zone in accreting white dwarfs. This process strongly modifies the downward mixing of the 
accreted matter and must be taken into account for determining the accretion rates. 
Here we present preliminary results obtained for the computation of thermohaline time scales in a DA white dwarf model, with  parameters representative of the star G29-38:
 $M_{*}/M_{\odot}$=0.59, $M_{H}/M_{*}$=5$\times$$10^{-10}$, $M_{He}/M_{*}$=2.5$\times$$10^{-2}$, $L/L_{\odot}$= 0.0026, $T_{eff}$= 11100 K as derived from asteroseismology 
(\citet{romero12}). These results can already lead to the conclusion that the effect is very important for DAZ and may increase the derived accretion rate by 
several orders of magnitude

\section{Consequences for accreting white dwarfs}

The accretion/mixing process described in section 2 is expected to occur in white dwarfs in the same way as in accreting main sequence stars.  When a DA or DB white dwarf 
accretes heavy matter from a debris disk, the same events as described in \citet{vauclair04} occur successively. The new chemical elements are mixed in the 
convective zone, but they do not stay inside this zone because of the induced inverse $\mu$-gradient.  First dynamical convection occurs below the Schwarzschild zone, as 
predicted by the ``Ledoux criterion'', until the $R_{0}$ ratio becomes equal to one. Then thermohaline convection begins, and the accreted matter is diluted far below the
 dynamical convective zone.     

For the considered DA model, we have computed the gravitational diffusion time scale for calcium, in the same way as in \citet{theado09}, and the 
thermohaline convection time scale, computed with the \citet{vauclair12} thermohaline mixing coefficient, as a function of the external mass, 
between the bottom of the convective zone and the H-He transition zone. The thermohaline time scale is computed as the time needed for the accreted matter to reach the 
considered layer. We find that the thermohaline time scale is several orders of magnitude smaller than the gravitational diffusion time scale, as expected.  The values
 below the convective zone may be as low as a few minutes for the thermohaline time scales, compared to a few years for the diffusion time scale. At the bottom of the 
hydrogen zone they become closer, of the order of one year for thermohaline, ten years for diffusion. This means that the accreted matter is rapidly mixed in all the 
hydrogen layer. Thermohaline mixing is completely stopped at the transition region, because of the $\mu$ increase induced by the increasing amount of helium. 

These preliminary computations prove that the accretion rates needed to account for the heavy element abundance observations are much larger than generally assumed.
 Reaching a present $\mu$ value of 0.5005 as observed in G29-38 would need an original $\mu$ value as large as 0.532. This is probably too large to correspond to a single 
accretion episode. Time dependent computations with several smaller accretion episodes are needed to go further. This is a work in progress.

\section{Conclusion}

We have shown that the accretion of heavy elements onto white dwarfs induces a thermohaline instability due to the created inverse $\mu$-gradient. The accreted material is
 mixed down in a much deeper part of the star than the usually considered convective zone. In our preliminary study of this effect in a representative DAZ model close to 
G29-38, the accreted heavy elements are diluted in a fractional mass of the star ~ $10^{3}$ larger than the mass of the convective zone. It follows that accretion rates 
should be $10^{3}$ larger to produce the observed abundances of heavy elements. The effects of thermohaline instabilities in DAZ with larger hydrogen mass fractions and 
in DBZ are under study. We expect the thermohaline convection to be less efficient in DB than in DA white dwarfs due to the different initial $\mu$ value and the different 
internal structure. In any case, we demonstrate that this special convection process has to be taken into account in all computations of the accretion rates needed to 
account for the observations. It will reduce and may even suppress the differences in the accretion rates needed to account for the observations of DAZ and DBZ stars. 
This will be presented in a forthcoming paper. 

\acknowledgements 
We thank Dr. Noel Dolez for providing the white dwarf model used in these computations.


\end{document}